\documentclass[epj]{svjour}
\usepackage{amssymb}
\usepackage{amsmath}
\usepackage{array} % to improve presentation of tables
\usepackage{epsfig}
\usepackage{graphics}
\usepackage{colortbl}
\usepackage{hhline}
\usepackage{multirow}
\begin{document}
\title{High-Energy Scattering vs Static QCD Strings}
%\subtitle{Do you have a subtitle?\\ If so, write it here}
\author{V.A.~Petrov\thanks{\emph{e-mail:} Vladimir.Petrov@cern.ch}\inst{1}, R.A.~Ryutin\thanks{\emph{e-mail:} Roman.Rioutine@cern.ch}\inst{1}
}                     % Do not remove
%
%\offprints{}          % Insert a name or remove this line
%
\institute{{\small Institute for High Energy Physics},{\small{\it 142 281}, Protvino, Russia}}
%
%\date{Received: date / Revised version: date}
% The correct dates will be entered by Springer
%
\abstract{We discuss the shape of the interaction region of the 
elastically scattered protons stipulated by the high-energy  
Pomeron exchange which turns out to be very similar with the 
shape of the static string representing the confining QCD flux tube.
This similarity disappears when we enter the LHC energy region, which
corresponds to many-Pomeron exchanges. Reversing the argument we conjecture
a modified relationship between the width and the length
of the confining string at very large lengths.
 % end of PACS codes
} %end of abstract
\authorrunning
\titlerunning
\maketitle
%

%----------------Introduction---------------------------------------------

\section{Itroduction}

The basic element of the Regge framework to describe the high-energy hadron scattering is the well-known expression 
\begin{equation}
\label{eq:ampregge}
A(s,t)=\beta(t)\left( \mathrm{e}^{-\frac{\mathrm{i}\pi}{2}}s\right)^{\alpha(t)}+\dots
\end{equation}
where $\alpha(t)$ is the leading (Pomeron) Regge trajectory. The 
average transverse ( w. r. t. the collision axis) size of the 
interaction region is given by the formula
\begin{equation}
\label{eq:averagetrsize}
<{\vec{b}}^2>=4\alpha^{\prime}(0)\ln(s)+\dots
\end{equation}
Thus, in the Regge-pole framework the transverse size of the interaction 
region widens with energy. 

It is not difficult to obtain the average longitudinal range of 
interaction which appears to be
\begin{eqnarray}
R&=&<z_1-z_2>=<\partial(scattering\;phase)/\partial p^{out}_{\parallel}>=
\nonumber\\
\label{eq:Lsize}&=&\pi
<\alpha^{\prime}(t)>\sqrt{s}/2+\dots
\end{eqnarray}
where $z_{1,2}$ are longitudinal coordinates of the scattered 
hadrons in the c.m.s., $scattering\;phase= -\pi\alpha(t)/2$, 
$t=-2p^2+ 2pp_{\parallel}^{out}$, $s=4p^2+4m^2$.

In  linear approximation for 
the trajectory, 
$\alpha(t)= \alpha(0)+\alpha^{\prime}(0)t$, we obtain
\begin{equation}
\label{eq:Rlinear}
R=\pi\alpha^{\prime}(0)\sqrt{s}/2+\dots
\end{equation}
The fast growth with energy of the longitudinal interaction range
in diffractive scattering is known for a long time (see f.e.~\cite{1,1a,1b,1c,1d,1e}).

We get the following relationship between 
transverse ($<{\vec{b}}^2>$) and longitudinal ($R$) sizes 
of elastic hadron scattering (at large $R$)
\begin{equation}
\label{eq:TRlinear}
<{\vec{b}}^2>=8\alpha^{\prime}(0)\ln(R)+\dots   
\end{equation}
So the transverse size of the interaction region increases 
logarithmically with the growth of the longitudinal interaction 
range. Pictorially at large $R$ it looks as a prolate cigar-like region:
\begin{figure}[h!]  
 \includegraphics[width=0.49\textwidth]{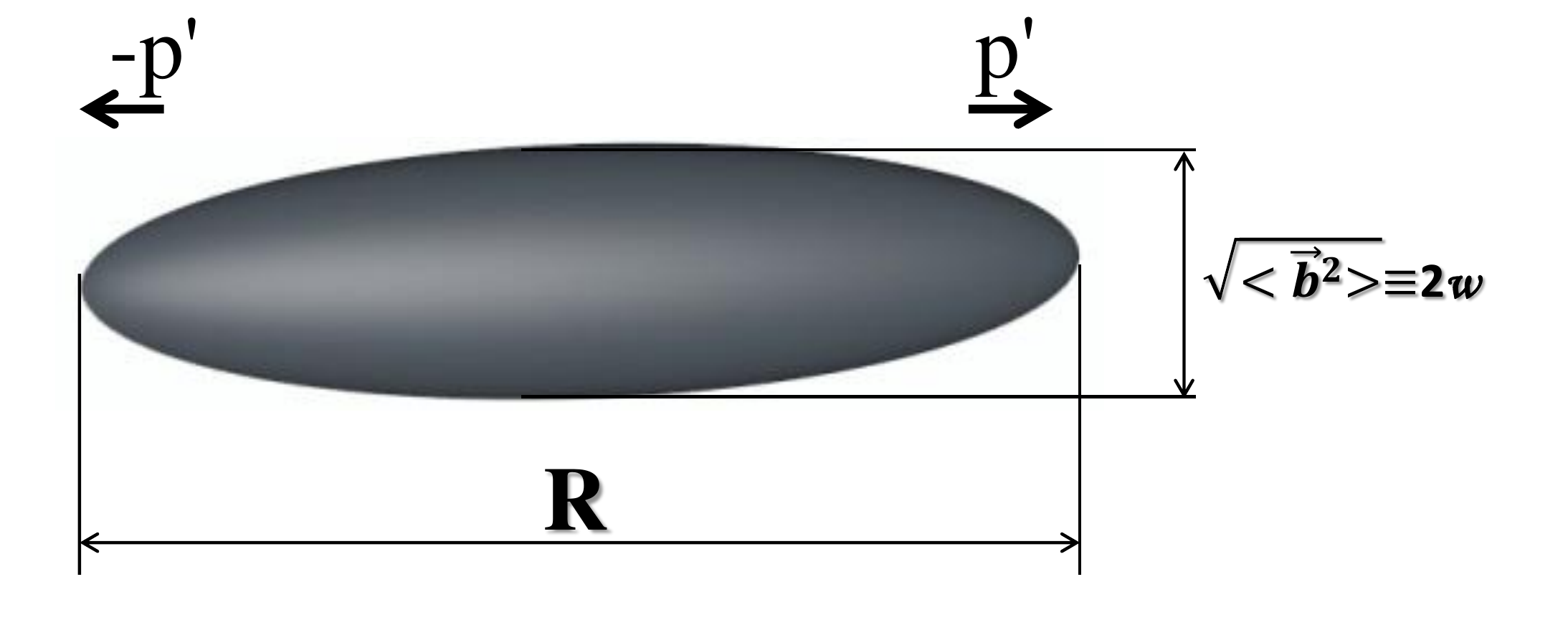} 
  \caption{\label{fig1:intreg} Interaction region.}
\end{figure}

\section{The Shape of Static Strings}
In paper ~\cite{luscher} the question of the thickness of a static string spanned by
infinitely heavy quark-anti-quark pair was addressed. Trying different frameworks the 
authors came to the conclusion that (the square of )the size of the transverse quantum fluctuations 
of a string-like confining flux grows logarithmically with the string length R.

This growth is thought to overcome the classical mechanism of the chromo-electric flux squeezing (dual Meissner effect).

Much later the same problem was studied in a more rigorous
way. In particular, in Ref.~\cite{1e,meyer} 
it was shown that in $d=2+1$ dimensional space-time the 
confining QCD string connecting two static sources 
($Q\bar{Q}$) has the square of half-width $w^2$ (in the middle
of the string length), which 
grows with the length of the string $R$ as  
\begin{equation}
\label{eq:w2cite}
w^2=\frac{d-2}{2\pi\sigma}\ln(R)+\dots
\end{equation}
where $\sigma$ is the string tension.

If we assume that at high energies the interaction region 
is close to axisymmetric form, then we have for its half-width squared
\begin{equation}
\label{eq:w2assume}
w^2_{scat}=\frac{<{\vec{b}}^2>}{4}=2\alpha^{\prime}(0)\ln(R)+\dots
\end{equation}
If to take into account the relation between the Regge slope and string tension
\begin{equation}
\label{eq:relsloptens}
\alpha^{\prime}=\frac{1}{2\pi\sigma}
\end{equation}
then~(\ref{eq:w2cite}) and~(\ref{eq:w2assume})
coincide if $d=3+1$, i.e. formally outside the scope of the 
derivation of (\ref{eq:w2cite}) in Ref.~\cite{1}, where $d=2+1$. 

However, in paper~\cite{2} the same problem was studied in 4-dimensional 
space. The authors of this paper used the same logarithmic relation with the coefficient $B$ in front of $\ln(R)$ as one of two free parameters 
for the fitting the lattice data. Below we show their 
picture.
\begin{figure}[h!]  
 \includegraphics[width=0.49\textwidth]{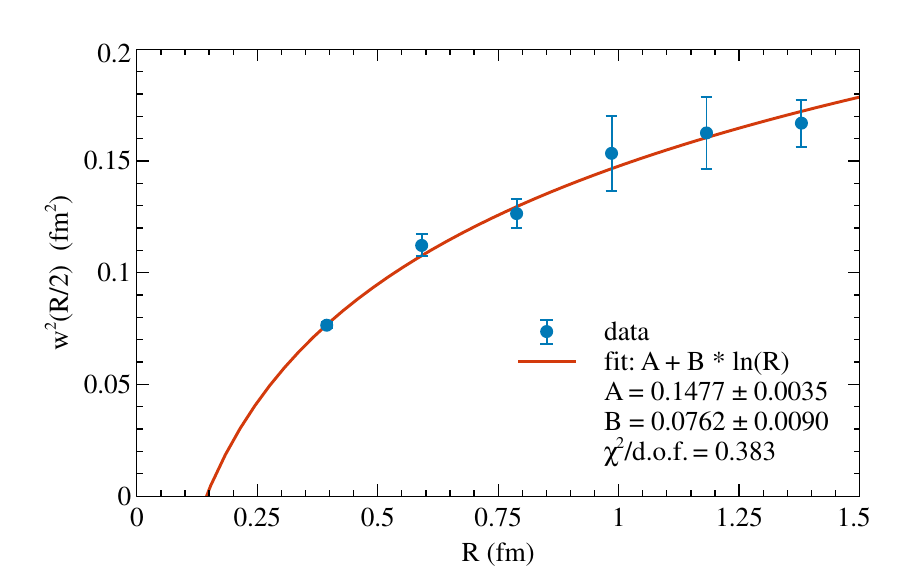} 
  \caption{\label{fig2:w2vsR} The width squared of the $q\bar{q}$ 
  confining string as function of its length~\cite{2}.}
\end{figure}
\section{Reggeon Exchanges and Strings}
If we identify the ``static'' coefficient 
$B=0.0762\pm0.0035\;\mathrm{fm}^2$ with $2\alpha^{\prime}(0)$ then we obtain 
\begin{equation}
\label{eq:aprime}
\alpha^{\prime}(0)=0.9525\pm 0.04375\;\mathrm{GeV}^{-2}
\end{equation}

This slope is not very far from the values of the secondary Regge 
trajectories which are usually associated with open $q\bar{q}$ 
strings. So if the high-energy scattering were defined by the 
meson trajectories we could conclude that the interaction region 
is being stretched  with the energy growth as an open 
string.

We, however, know that actually it is not the meson (secondary Reggeon) but 
the Pomeron trajectory which dominates the 
high-energy behaviour. In this case the coefficient in 
front of  $\ln(R)$ is four times smaller. How can it be explained?

Usually one argues that taking into account the closed string nature 
of the Pomeron we should get a twice smaller coefficient 
corresponding to $\alpha^{\prime}_{{\mathbb{P}}}=0.5\;\mathrm{GeV}^{-2}$. However 
the high-energy phenomenology supports in a very strong way 
even smaller slope:
\begin{equation}
\label{eq:Pomaprime}
\alpha^{\prime}_{{\mathbb{P}}}\cong 0.25\;\mathrm{GeV}^{-2}
\end{equation}

If we assume that the closed (stretched) string 
shape follows Eq.(\ref{eq:w2cite}) with 4 times smaller coefficient 
in front of $\ln(R)$ then we obtain the following estimate for the Pomeron slope: 
\begin{equation}
\label{eq:PomaprimeRef2}
\alpha^{\prime}_{{\mathbb{P}}}(0)\simeq 0.24\pm0.01\;\mathrm{GeV}^{-2}.
\end{equation}
Actually the tension values of various kinds of closed strings 
in the lattice framework were considered in Ref.~\cite{3} 
with one of the results fairly close to that of 
high-energy phenomenology:
\begin{equation}
\label{eq:PomaprimeRef3}
\alpha^{\prime}_{{\mathbb{P}}}(lattice)\simeq 0.253(2)\;\mathrm{GeV}^{-2}.
\end{equation}

This is suggestive for the conclusion that the interaction region 
of high-energy hadron scattering  clearly bears stringy features 
similar to the static string configurations in spite of their 
physically extraordinary difference: a highly dynamic 
phenomenon against the static one. In the case of
scattering we deal with fluctuating average length
while in Eq.~(\ref{eq:w2cite}) it is fixed parameter.

\section{Breakdown of Reggeon-String Similarity or String Shape Modification?}
We have to notice that Regge-Pomeron formula~(\ref{eq:w2cite}) begins 
to dominate at energies not less than the ISR 
energies ($\sqrt{s}=20\div50\;\mathrm{GeV}$), this corresponds to 
the minimal interaction length $R=1.5$~fm, which is exactly at the 
end of the lattice data at Fig.\ref{fig2:w2vsR}.

%Moreover at presently available energies at the LHC the 
%(average value of) the longitudinal interaction range $R$ achieves 
%huge values of order of $2500$~fm. It is also worth to mention 
%that this average value is subject to enormous fluctuations 
%which amount to $\Delta R_{min}\approx 14000$~fm at $7$~TeV. 
If we believe 
that the interaction region looks like a stretched string of non-zero 
transverse extent  then we come to a striking conclusion that 
these extremely stretched configurations can survive without 
breaking up to monstruous lengths of order of $150$~fm. This concerns
high-mass ($\simeq 1\; \mathrm{TeV}/c^2$) stringy configurations.

According to the ratio $\sigma_{el}/\sigma_{tot}$ measured by the 
TOTEM LHC Collaboration~\cite{4}
\begin{equation}
\label{eq:totemeltot}
\sigma_{el}/\sigma_{tot}(\mathrm{TOTEM},\;7\;\mathrm{TeV})=0.257\pm0.005
\end{equation}
this means that the probability of the long string-like configuration  
survival (before breaking out into the two elastically 
scattered protons) is of order of 25\%. Such a phenomenon would 
mean that confinement is fairly important not only at 
low-energy hadron physics but at high-energy scattering as well.

Our arguments are heavily based on Eq.(\ref{eq:w2assume}). However it 
ceases to be adequate at very high energy though remains a 
basic building block (eikonal) for the full unitary amplitude in 
the Regge-eikonal framework. Latest experiments at the LHC show 
that the width $w^2$ as function of $R$ significantly deviates 
from the simple logarithmic dependence in the LHC 
energy region (see Fig.~\ref{fig3:expw2vsR} below).

\begin{figure}[h!]  
 \includegraphics[width=0.49\textwidth]{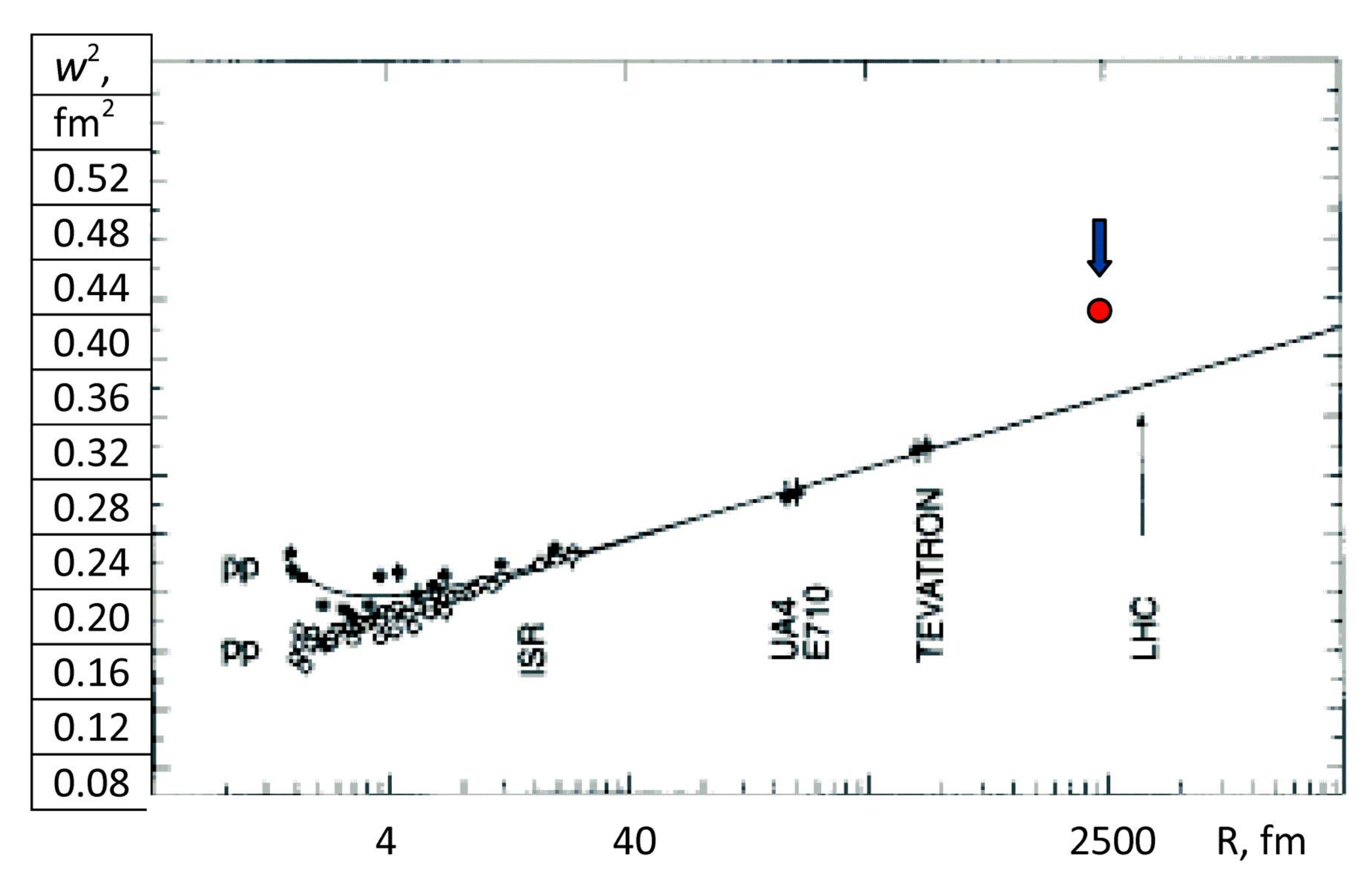} 
  \caption{\label{fig3:expw2vsR} The length dependence of the 
  half-width of the transverse interaction region of the 
  elastic scattering of (anti)protons. $w^2_{scat}=\frac{<{\vec{b}}^2>}{4} 
  \approx 2\alpha^{\prime}_{{\mathbb{P}}}(0)\ln\left(R\right)$
  $\approx (0.14\;\mathrm{fm})^2\ln\left(R\right)$. The 
  red point is 
  the measurement of the TOTEM Collaboration at $7$~TeV. The data 
  from the ALFA experiment (ATLAS)~\cite{5} gives a 
  close value. (The figure is a modification of Fig.35 from 
  Ref.~\cite{5})}
\end{figure}

At ultra-high energies the relationship between $w^2$ and $R$ 
changes significantly as could be seen from the 
following  asymptotic behavior:
\begin{eqnarray}
&&<{\vec{b}}^2>(ultra-high\;energy)\approx\nonumber\\
&&\label{eq:b2as}\phantom{<{\vec{b}}^2>} 4\left[
\alpha_{{\mathbb{P}}}(0)-1
\right]
\alpha^{\prime}_{{\mathbb{P}}}(0)
\left[\ln (s)\right]^2+\dots
\end{eqnarray}
Unfortunately we haven't yet the analytic expression for the 
phase of the scattering amplitude in this case and can't 
insist that the proportionality 
$$
R\sim \sqrt{s}
$$
still holds. However to trace the trend we can use the asymptotic 
(actually at practically unavailable energies)
expression for the scattering amplitude:
\begin{eqnarray}
\label{eqq:Tst} T(s,t)&\approx& \frac{4\pi\mathrm{i} s R_0(s)}{\sqrt{-t}} J_1 (R_0(s)\sqrt{-t}),\\
  R_0^2(s)&=&4\Delta_{{\mathbb P}}\alpha^{\prime}_{{\mathbb P}}(0)
\left[\ln(s)-\mathrm{i}\pi/2\right]^2,\nonumber\\
\label{eqq:R0} \Delta_{{\mathbb P}}&=& \alpha_{{\mathbb P}}(0)-1.
\end{eqnarray}
We get for the interaction region scales:
\begin{eqnarray}
\label{eqq:R} R&\approx&\pi\alpha^{\prime}_{{\mathbb P}}(0)\Delta_{{\mathbb P}}\sqrt{s}\ln(s)+\dots,\\
\label{eqq:b2} <\vec{b}^2>&=&4\alpha^{\prime}_{{\mathbb P}}(0)\Delta_{{\mathbb P}} 
\left[ \ln(s)\right]^2+\dots.
\end{eqnarray}
If we try to assume that the similarity discussed above
persist at higher energies (larger $R$) then we come to the
following modification of the string width-length relationship: 
\begin{equation}
\label{eqq:w2}w^2_{string} =2\frac{\Delta_{{\mathbb P}}}{\pi\sigma_{{\mathbb P}}}
\left[ \ln(R)\right]^2+\dots,
\end{equation}
where $\sigma_{{\mathbb P}}$ is the tension of the string corresponding to the Pomeron while parameter $\Delta_{{\mathbb P}}$ can be only explained when considering the trajectory deviation from the linear behavior at small unphysical values of the string mass (see e.g. Ref.~\cite{X}). We have also to note that the value of the tension $\sigma_{{\mathbb P}}=1/(2\pi\alpha^{\prime}_{{\mathbb P}}(0))$ may differ from the ``quasi-classical'' tension of the straight-line  string at high string masses.

\begin{figure}[h!]  
 \includegraphics[width=0.49\textwidth]{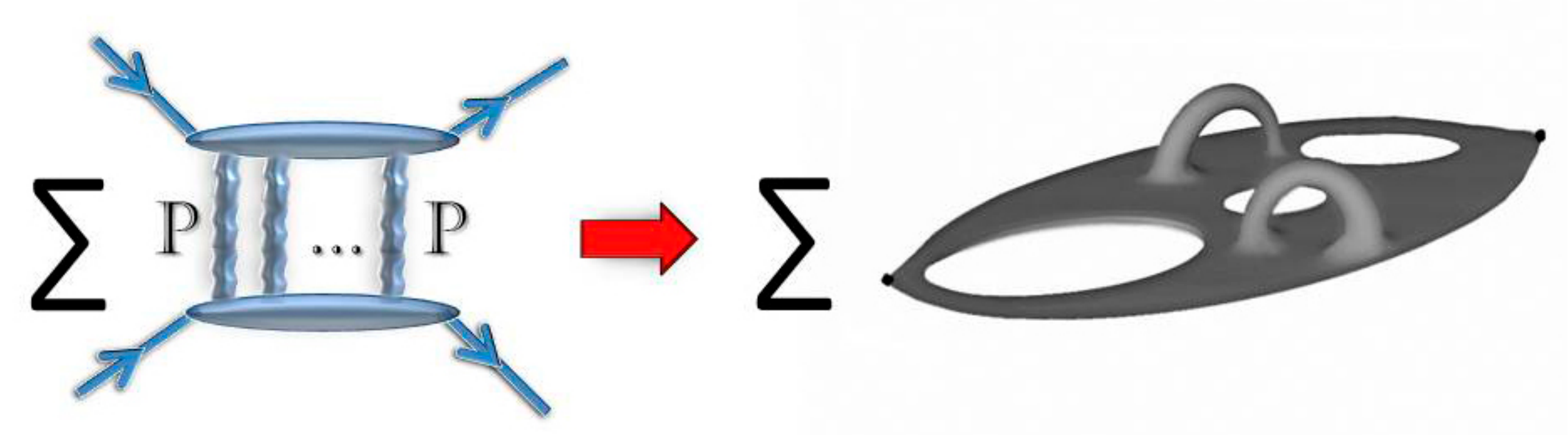} 
  \caption{\label{fig4} Towards the similarity of multi-Reggeon exchanges and account of possible topologies of the string.}
\end{figure}

The asymptotic form of the scattering amplitude (\ref{eqq:Tst}) is due to summation of infinitely many Reggeons. If 
to liken the Reggeon to a string (a closed one in case of the Pomeron) then this would correspond to summation 
in all possible genera (handles and holes) of the string spanned between the 
sources (see Fig.~\ref{fig4}). We could conjecture that account of 
higher genera in the problem of the static string width dependent on its length leads to formula (\ref{eqq:w2}).

\section{Conclusions}

 We have shown that the relationship of the transverse and longitudinal 
 interaction ranges which follow from a one-Reggeon-exchange scattering amplitude is identical to that of the static confining string.
 Reversing the argument we conjecture on the basis of the Regge-eikonal asymptotic for the scattering amplitude that account of higher genera modify the dependence of the static string width on its length.

\section*{Aknowledgements}

We would like to thank our colleagues A.A.~Godizov, \linebreak A.K.~Likhoded
and G.P.~Pronko for discussions.

%
% BibTeX users please use
% \bibliographystyle{}
% \bibliography{}
%
% Non-BibTeX users please use

\end{document}